\magnification=1200
\baselineskip=18truept

\line{\hfill RU-98-03}

\vskip 2truecm
\centerline{\bf More about exactly massless quarks on the lattice.}

\vskip 1truecm
\centerline{Herbert Neuberger}
\vskip .5truecm

\centerline {Department of Physics and Astronomy}
\centerline {Rutgers University, Piscataway, NJ 08855-0849}

\vskip 1.5truecm

\centerline{\bf Abstract}
\vskip 0.75truecm
In a previous publication [hep-lat/9707022] I showed that the fermion
determinant for strictly massless quarks can be written on the lattice
as $\det D$, where $D$ is a certain finite square matrix explicitly 
constructed from the lattice gauge fields. Here I show that $D$
obeys the Ginsparg-Wilson relation $D\gamma_5 D = D\gamma_5 +\gamma_5 D$.

\vfill\eject
In a recent publication [1] I showed that the overlap led to a simple
definition of a lattice gauge theory with exactly massless quarks on
the lattice. The vector-like character of the theory makes it possible
to represent the lattice Dirac operator for strictly massless
quarks by a matrix $D$ of finite size and fixed shape. 

$D$ is defined as follows: Start from,
$$
X(m) = \pmatrix {B+m & C\cr -C^\dagger & B+m },\eqno{(1)}$$
where,
$$
\eqalign{
( C )_{x \alpha i, y \beta j}
& ={1\over 2} \sum_{\mu=1}^{4} \sigma_\mu^{\alpha\beta}
[\delta_{y,x+\hat\mu} (U_\mu (x) )_{ij} -
\delta_{x,y+\hat\mu} (U_\mu^\dagger (y))_{ij}] ,\cr
( B )_{x \alpha i, y \beta j} & = {1\over 2} \delta_{\alpha\beta} 
\sum_{\mu =1}^{4} [2\delta_{xy}
\delta_{ij} - \delta_{y,x+\hat\mu} (U_\mu (x) )_{ij} -
\delta_{x,y+\hat\mu} (U_\mu^\dagger (y))_{ij}],\cr}\eqno{(2)}$$
and 
$$
\pmatrix {0 & \sigma_\mu \cr \sigma_\mu^\dagger & 0} =\gamma_\mu . 
\eqno{(3)}$$
The $\gamma_\mu$ are Euclidean Dirac matrices, $x,y$ are sites on 
the lattice, $\alpha ,\beta$ are Weyl spinor indices and $i,j$ 
are color indices. The $U_\mu (x)$ are lattice link matrices. Set
the parameter $m$ to some number in the range $(-1,0)$. Define the
unitary matrix $V$ by
$$
V=X{1\over {\sqrt {X^\dagger X }}}.\eqno{(4)}$$
This definition is valid except for exceptional 
configurations with $\det (X)=0$. We shall assume that these gauge backgrounds
can be ignored statistically. Now define $D=1+V$.

In ref. [1] I first argued that $\det D$ was a good 
lattice regularization of the fermion determinant for exactly massless
quarks, and then showed that $D$ represented the effects of instantons 
correctly, by robust zeros. Other nice features of the $D$
were also pointed out. 
Obviously, the spectrum of $D$ 
is concentrated on the circle $1+e^{i\theta},~ \theta \in [0,2\pi]$ 
in the complex plane. In odd Euclidean dimensions
this property also holds and is instrumental in checking
that the correct global anomalies are reproduced on the lattice [2].

Here (see also [3]) 
I wish to add the rather trivial observation that, in view of
$$
\gamma_5 X \gamma_5 = X^\dagger ,\eqno{(5)}$$
we also have 
$$
\gamma_5 {1\over{1+V}} \gamma_5 = 
{1\over {1+\gamma_5 V \gamma_5 }}= {1\over {1+V^\dagger }} =
{V\over{1+V}} = 1- {1\over {1+V}}.\eqno{(6)}$$
This means
$$
\gamma_5 D^{-1} +D^{-1} \gamma_5 = \gamma_5 ,\eqno{(7)}$$
an equation Ginsparg and Wilson first wrote down many years ago [4],
as a way one may represent exact masslessness on the lattice, while,
at the same time, preserving the continuum anomaly. In ref. [4]
eq. (7) (actually a slight generalization of equation (7) which
is immaterial here) was derived as a ``remnant of chiral symmetry
on the lattice'' after blocking a chirally symmetric
continuum theory with a necessarily chirality breaking 
local renormalization group kernel, of the type studied thoroughly 
in ref. [5] for example. 

As shown in ref. [1], $D$ nicely reproduces instanton effects. 
Once it is understood that $D$ can be used to define topological
charge, since the latter is an integer valued, nonconstant
function over the compact space of gauge configurations (we are
assuming a finite lattice size), we know that exceptional 
configurations invalidating some of the definitions must
exist. We identified them above as those configurations
for which $X$ becomes non-invertible. Thus, what we had to
designate as an ``exceptional'' configuration turns out to be
very close to the definition 
one adopts in lattice QCD with``ordinary'' Wilson fermions.
The topological charge defined
from $D$ has been shown to produce reasonable quantitative
and qualitative results in [6].

A recent paper [7] presents a very implicit definition of another
matrix $D$ which also obeys eq. (7), and has a similar spectrum
in its simplest variant. Strictly speaking, this matrix is infinite,
but probably admits some truncations that would have no discernible
effect numerically. 
The matrix $D$ in this paper 
and the related methods of extracting a topological charge 
from lattice gauge field configurations mentioned above, 
have been arrived at in the overlap framework, 
independently of ref. [7]. On the other hand, 
the authors of ref. [7] appear
oblivious of the overlap. It is therefore
a quite amusing coincidence that relatively
similar solutions to the problem of putting strictly massless
quarks on the lattice have been arrived at from quite different
starting points, independently. The result of ref. [7] provides
further support to the overlap, although it is unclear whether
any such support is still needed, given the impressive
numerical results of [8]. (Actually, since full implementation
of the matrix $D$ on the lattice would be expensive due to the
square root factor, one needs to truncate, essentially approximating
the overlap. The truncation is studied in refs. [9], but
has been used in [8] before [9] appeared, since it was proposed before, in 
[10], as an improvement over [11].)

The most obvious differences between
the two ways of defining a $D$ matrix is that the one given here
and in ref. [1] is explicit. The matrix 
elements of the $D$-matrix of ref. [7]
are defined implicitly as the solution of a nontrivial
recursion relation, which, in turn, includes internally a nontrivial
minimization in the space of gauge fields. 
In practice
it would seem easier to use the overlap (via the flow methods of [6]) 
for getting the topological charge. For many more uses of $D$ 
(more precisely, its truncation) I refer again to [8].

The claim of reference [7] about the absolute 
absence of exceptional configurations, 
in the sense that I use the term, cannot
hold, since they are a logical necessity for
any definition of topological charge on the lattice, as argued 
above. Let us consider the vicinity of an exceptional gauge configuration:
Changing the background ever so slightly we can get topological
charge zero or one and it is quite plausible that some of the zero
modes we find are better viewed as lattice artifacts.
The space of all continuum connections over
a compact manifold is not connected while the
replacement of this space in the lattice approximation to the manifold
clearly is. There is no way around this and and the price
to pay will always be in accepting the presence of 
some exceptional configurations.

I should add a word of caution here: Clearly, identity (7)
would hold even had we picked the parameter $m$ in (1) positive.
This would eliminate all exceptional configurations since one
can easily prove (see first paper in [6]) that $\det X \ne 0$ for
any gauge field. However, the new matrix $D$ does not describe
massless quarks, and, if ``asked'' what the topology of the gauge
background is, would always return zero for an answer. 

{\bf Acknowledgment:} This research was 
supported in part by the DOE under grant \#
DE-FG05-96ER40559. 
\medskip

{\bf  References:}

\smallskip

\item{[1]} H. Neuberger, hep-lat/9707022, Phys. Lett. B, to appear.
\item{[2]} Y. Kikuakwa, H. Neuberger, hep-lat/9707022, Nucl. Phys. B, to
appear.
\item{[3]} H. Neuberger, hep-lat/9710089.
\item{[4]} P. Ginsparg, K. Wilson, CLNS-81/520, HUTP-81/A060, Dec. 1981. 
\item{[5]} T. Balaban, M. O'Carroll, R. Schor, Lett. Math. Phys. 17 (1989) 209.
\item{[6]} R. Narayanan, H. Neuberger, Phys. Rev. Lett. 71 (1993) 3251,
Nucl. Phys. B443 (1995) 305; R. Narayanan, P. Vranas, hep-lat/9702005,
R. G. Edwards, U. M. Heller, R. Narayanan, hep-lat/9801015.
\item{[7]} P. Hasenfratz, V. Laliena, F. Niedermayer, hep-lat/9801021. 
\item{[8]} T. Blum, A. Soni, Phys. Rev. D56 (1997) 174;
Phys. Rev. Lett. 79 (1997) 3595.
\item{[9]} H. Neuberger, hep-lat/9710089; 
Y. Kikuakwa, H. Neuberger, A. Yamada, \hfill\break hep-lat/9712022.
\item{[10]} Y. Shamir, Nucl. Phys. B406 (1993) 90.
\item{[11]} D. B. Kaplan, Phys. Lett. B288 (1992) 342;
R. Narayanan, H. Neuberger, Phys. Lett. B302 (1993) 62.

\vfill\eject\end